\documentclass[twocolumn,aps,prb,epsf]{revtex4}
\usepackage[dvips]{graphics}
\begin{document}
\draft


\title{Theory of induced quadrupolar order 
in tetragonal YbRu$_{2}$Ge$_{2}$}

\author{Tetsuya Takimoto and Peter Thalmeier}

\address{Max Planck Institute for Chemical Physics of Solids, 
N$\ddot{o}$thnitzer Str. 40, 01187 Dresden, Germany}

\date{\today}

\begin{abstract}
The tetragonal compound YbRu$_{2}$Ge$_{2}$ exhibits a non-magnetic
transition at  $T_0$=10.2K  and a magnetic transition at $T_1$=6.5K in zero magnetic field. We present a model for this material based on a quasi-quartet of Yb$^{3+}$ crystalline electric field (CEF) states  and discuss its mean field solution.
Taking into account the broadening of the specific heat jump at $T_0$ 
for magnetic field perpendicular to [001] and the decrease of  $T_0$ 
with magnetic field parallel to [001], it is shown that 
ferro-quadrupole order of either O$_{2}^{2}$ or O$_{\rm xy}$ - type
are prime candidates for the non-magnetic transition. 
Considering the matrix element of these quadrupole moments, 
we show that the lower CEF states of the
level scheme consist of a $\Gamma_{6}$ and a $\Gamma_{7}$ doublet.
This leads to induced type of O$_{2}^{2}$ and O$_{\rm xy}$ 
quadrupolar order parameters. 
The quadrupolar order introduces exchange anisotropy 
for planar magnetic moments. 
This causes a spin flop transition at low fields perpendicular [001] 
which explains the observed metamagnetism.
We also obtain a good explanation for the temperature dependence of magnetic susceptibility and specific heat for fields both parallel and perpendicular to the [001] direction.
\end{abstract}

\maketitle
\narrowtext

\section{Introduction}
It is well known that some $f$-electron systems show multipole ordering. 
The phenomenon has attracted much attention, since features of multipole 
order are quite different from usual magnetic order. 
As a typical case, CeB$_{6}$ shows a kind of 
antiferro-quadrupole ordering at 3.4K, whose transition temperature 
increases with increasing magnetic field\cite{Fujita,Takigawa,Effantin}. 
For transition in NpO$_{2}$, 
an octupole ordering is proposed 
due to experimental results of resonant x-ray scattering 
\cite{Mannix,Paixao,Lovesey} 
and NMR \cite{Tokunaga} 
although a cusp at the transition temperature 
is observed in uniform susceptibility\cite{Ross,Erdos}. 
There are at least two common properties between these compounds. 
At first, these compounds have cubic crystal structure. 
Secondly, the crystalline electric field (CEF) ground 
states of corresponding 
level schemes are quartet states, which are available only in cubic systems. 
It is thought that the 
quartet state is responsible for multipole ordering. 

Recently, some anomalous properties have been observed 
in the tetragonal metallic compound 
YbRu$_{2}$Ge$_{2}$\cite{Jeevan}. 
In specific heat measurement without magnetic field, there are three 
transition temperatures at $T_0$=10.2K, $T_1$=6.5K, and $T_2$=5.7K. 
It is important that the entropy around $T_0$ obtained by integration 
of specific heat data is very close to Rln4, which means the existence of a
quasi-quartet state even in the tetragonal system. 
Applying a magnetic field perpendicular to [001] direction, 
the specific heat jump at $T_0$ broadens, and the peak position 
corresponding to 
$T_0$ seems to increase, while $T_1$ and $T_2$ merge and decrease. 
Increasing magnetic field further above 7T, no anomaly is found. 
On the other hand, in magnetic field parallel to [001], 
not only $T_1$ and $T_2$ but also $T_0$ decrease 
with increasing magnetic field. 
For the magnetic susceptibility $\chi_{\rm ab}$ in magnetic field 
perpendicular to [001], no anomaly appears at $T_0$, while a cusp 
is observed at $T_1$ for small magnetic fields. 
Furthermore, a metamagnetic transition at higher fields around 2T 
is regarded as spin-flop transition. 
The magnetic susceptibility $\chi_{\rm c}$ in field 
parallel to [001] is almost temperature independent between 
$T_0$ and $T_1$, and shows flat temperature dependence below $T_1$ 
after a slight decrease just below $T_1$. 
Because the value of the paramagnetic effective moment 
4.5$\mu_{\rm B}$ is 
very close to magnetic moment of free Yb$^{3+}$, 
it is a reasonable assumption that 
$f$-hole of Yb$^{3+}$ is almost localized. 
Considering the quasi-quartet state in a localized picture, 
some multipole moments will be active at each site 
in the system. 
From these experimental data, it has been suggested that $T_0$ is 
a kind of quadrupole transition, 
while the phase below $T_1$ is regarded as 
antiferromagnetic phase with planar staggered moment\cite{Jeevan}. 
According to Jeevan ${\it et}$ ${\it al.}$, a change in magnetic 
structure may happen at $T_2$. We will ignore this subtlety 
in the following and consider only $T_1$. 

In the theoretical analysis of CeB$_{6}$, Shiina ${\it et}$ ${\it al.}$ 
have provided 
a mean-field approximation for the effective Hamiltonian of localized 
$f$-electrons belonging to $\Gamma_{8}$ irreducible representation 
in O$_{\rm h}$ point group.
In this case all multipole moments up to octupole are active \cite{Shiina}. 
The relevant multipoles have been classified according to 
irreducible representations of 
the point group in zero magnetic field. 
Taking into account that some symmetry operations of O$_{\rm h}$ point 
group elements are lost in a magnetic field, the multipoles 
have been reclassified according to irreducible representations 
of the relevant point group 
in the magnetic field. 
Using these multipoles 
a mean-field approximation has been applied to an 
effective Hamiltonian to construct the $H$-$T$ phase diagram. 
After introduction of anisotropic interaction of quadrupoles, 
they have obtained a consistent explanation 
for the anomalous ordering in CeB$_{6}$. 
It should be noted that this approach is promising 
for multipole ordering not only in CeB$_{6}$ but also 
in TmTe\cite{TmTe}, where Tm$^{2+}$ has 
the same (4{\it f})$^{13}$ electronic configuration 
as Yb$^{3+}$, and NpO$_2$\cite{Kubo}. 

In the present work, we apply this approach 
to investigate the phases of YbRu$_{2}$Ge$_{2}$. 
In this system, there are some significant differences to CeB$_{6}$, 
though a kind of quadrupole ordering is expected. 
At first, the crystal structure of YbRu$_{2}$Ge$_{2}$ is tetragonal, 
with point group D$_{\rm 4h}$. 
Second, noting that composition of the quasi-quartet depends on 
crystalline electric field parameters, 
it is expected that 
multipole ordering is also affected by the level scheme, 
while only the size of coupling constants decides on the 
favorable multipole ordering in cubic systems\cite{Shiina}. 
Third, considering the present system is tetragonal, 
some multipoles are described only by {\it induced moments}, 
whose expectation value in the CEF ground state vanishes
\cite{Trammell,Bleaney}. 
Especially, the second and third points bring additional complexity 
to identify a multipole transition. 
In order to explain the behavior of YbRu$_{2}$Ge$_{2}$, 
we introduce an effective Hamiltonian in the next section. 
Then, we apply a mean-field approximation for the Hamiltonian 
to identify the non-magnetic ordering state below $T_{0}$. 
Furthermore, we try to reproduce temperature dependences of 
specific heat and uniform susceptibility in magnetic field 
with anisotropic magnetic interaction. 
Finally, we summarize our results.

\section{Effective Hamiltonian}
From analysis of uniform susceptibility, effective 
moment is estimated as 4.5$\mu_{\rm B}$, which is quite close to 
the value 4.54$\mu_{\rm B}$ for free Yb$^{3+}$ ions. 
This means that the picture of localized hole in the 4$f$-shell 
will be reasonable for YbRu$_2$Ge$_2$. 
First we need to construct CEF level scheme of Yb$^{3+}$, 
to extract relevant multipole moments, 
and then we introduce effective intersite interactions 
between the multipole moments. 

\subsection{CEF term}
The total angular momentum $j$ of Yb$^{3+}$-ion is $j=7/2$. 
The multiplet splits into four Kramers-doublets in tetragonal 
crystal structure of YbRu$_2$Ge$_2$. 
In tetragonal point group D$_{\rm 4h}$, $j=7/2$ multiplet is classified 
according to two-$\Gamma_6$ and two-$\Gamma_7$ irreducible representations. 
For two doublets belonging to the same 
$\Gamma$-irreducible representation, 
we call the lower and higher ones $\Gamma^{(1)}$ and $\Gamma^{(2)}$, 
respectively, in the following. These states are described by linear 
combination of free ion states 
$|\mu\rangle$=$|j\mu\rangle$ ($|\mu|\leq\frac{7}{2}$) as follows, 
\begin{eqnarray}
  &&|\tau=1,\pm\rangle=
    |\Gamma_{6}^{(1)}\pm\rangle=\alpha_{11}|\frac{\pm 7}{2}\rangle
                             +\alpha_{12}|\frac{\mp 1}{2}\rangle,\\
  &&|\tau=2,\pm\rangle=
    |\Gamma_{6}^{(2)}\pm\rangle=\alpha_{21}|\frac{\pm 7}{2}\rangle
                             +\alpha_{22}|\frac{\mp 1}{2}\rangle,\\
  &&|\tau=3,\pm\rangle=
    |\Gamma_{7}^{(1)}\pm\rangle=\beta_{11}|\frac{\mp 5}{2}\rangle
                             +\beta_{12}|\frac{\pm 3}{2}\rangle,\\
  &&|\tau=4,\pm\rangle=
    |\Gamma_{7}^{(2)}\pm\rangle=\beta_{21}|\frac{\mp 5}{2}\rangle
                             +\beta_{22}|\frac{\pm 3}{2}\rangle,
\end{eqnarray}
where $\mu$ is z-component of total angular momentum 
and + ($-$) of left-hand side shows pseudo-spin up (down) 
in Kramers-doublets. 

Usually, CEF parameters are estimated by fitting calculated 
uniform susceptibility to the observed one. 
In addition, the inelastic neutron scattering (INS) gives 
important informations like splitting energy between the ground 
and first excited states in the level scheme. 
In recent INS experiment in YbRu$_2$Ge$_2$, the level scheme 
is reported with the splitting energy of 0.9 meV\cite{Geibel1}. 
Using the splitting energy, we have carried out 
the fitting of uniform susceptibility. 
Unfortunately, we could not obtain unique CEF level scheme 
from this procedure. 
However, from reasonable CEF level schemes, 
we obtain the following common features: 
(1) The splitting energy 
between the ground and first excited states is about 12K, 
which is estimated by reproducing the entropy 
obtained from specific heat data; 
(2) the ground and first excited states 
consist of one $\Gamma_6$ and one $\Gamma_7$ states, 
neither two $\Gamma_6$ nor two $\Gamma_7$ states; 
and (3) energy splittings of the second excited state 
from the ground state are at least thirty times larger than 
the observed transition temperature $T_0$=10.2K and 
the splitting energy between the ground and first excited states. 
Due to the third point, we can neglect upper two doublets, 
if we consider only low temperature region. 
The relevant CEF Hamiltonian is then given by
\begin{eqnarray}
  &&H_{\rm CEF}=-\sum_{{\bf i},\eta}\frac{\Delta_0}{2}
                (f_{{\bf i}6\eta}^{\dagger}f_{{\bf i}6\eta}
                -f_{{\bf i}7\eta}^{\dagger}f_{{\bf i}7\eta}),
\end{eqnarray}
where $\Delta_0$ is splitting energy from $\Gamma_6^{(1)}$ state 
to $\Gamma_7^{(1)}$ state. 
Here, $f_{{\bf i}\tau\eta}^{\dagger}$ is creation operator of 
$f$-hole with pseudo-spin $\eta$ 
in $\Gamma_{\tau}^{(1)}$ Kramers-doublet at site ${\bf i}$. 
Although we assume that the lower two doublets consist of 
one $\Gamma_6$ and one $\Gamma_7$ states, this assumption will be 
justified during identification of non-magnetic ordered state 
in YbRu$_2$Ge$_2$.

\subsection{Zeeman term}
The Zeeman term due to the applied field {\bf h} is given by
\begin{eqnarray}
  H_{\rm Z}=-g_{J}\mu_{\rm B}\sum_{\bf i}{\bf h}\cdot{\bf J}_{\bf i},
\end{eqnarray}
where ${\bf J}$, $g_{J}$, and $\mu_{\rm B}$ are total angular momentum, 
Land$\acute{\rm e}$ $g$-factor of Yb$^{3+}$, 
and Bohr magneton, respectively. 
With use of second quantization, x- and z-components of the angular momentum 
in $\Gamma_6$-$\Gamma_7$ subspace are expressed as
\begin{eqnarray}
  &&J^{\rm z}=c^{\rm z}_{66}S^{\rm z}_{66}+c^{\rm z}_{77}S^{\rm z}_{77},
    \label{jz}\\
  &&J^{\rm x}=c^{\rm x}_{66}S^{\rm x}_{66}+c^{\rm x}_{77}S^{\rm x}_{77}
             +c^{\rm x}_{67}\frac{1}{\sqrt{2}}(S^{\rm x}_{67}+S^{\rm x}_{76}),
    \label{jx}
\end{eqnarray}
with $\alpha$-component of pseudo-spin operator given by
\begin{eqnarray}
  S^{\alpha}_{\tau\tau'}=\frac{1}{2}\sum_{\eta,\eta'}
     f_{\tau\eta}^{\dagger}\sigma^{\alpha}_{\eta\eta'}f_{\tau'\eta'},
\end{eqnarray}
where ${\bf \sigma}^{\alpha}$ is $\alpha$-component of Pauli matrix. 
The coefficients $c^{\alpha}_{\tau\tau'}$ 
are expressed by $\alpha_{12}$ and $\beta_{12}$ as
\begin{eqnarray}
  &&c^{\rm z}_{66}=7-8\alpha_{12}^{2},\hspace{5mm}
  c^{\rm z}_{77}=-5+8\beta_{12}^{2},\\
  &&c^{\rm x}_{66}=4\alpha_{12}^{2},\hspace{5mm}
  c^{\rm x}_{77}=4\sqrt{3}\beta_{12}\sqrt{1-\beta_{12}^{2}},\nonumber\\
  &&c^{\rm x}_{67}=\sqrt{7}\sqrt{(1-\alpha_{12}^{2})(1-\beta_{12}^{2})}
                +\sqrt{30}\alpha_{12}\beta_{12}.
\end{eqnarray}
Therefore the coefficients 
$\alpha_{12}$ and $\beta_{12}$ which determine the
structure of $\Gamma_6^{(1)}$ and $\Gamma_7^{(1)}$ states 
are incorporated through anisotropic effective $g$-factors in Eqs.~(\ref{jz},\ref{jx}). 
We comment on the limiting case of $\alpha_{12}=\beta_{12}=1$, 
which are very close to values estimated by fitting of uniform 
susceptibility in III. B.. As we have mentioned above, we consider only 
quasi-quartet consisting of one $\Gamma_6$ and one $\Gamma_7$ doublet
at each site. As far as the quasi-quartet is concerned, multipole moments 
up to octupole are relevant as given in Table I. In this sense, 
the inter-site term will be always mapped to $j$=3/2 quartet system. 
In particular, with $\alpha_{12}=\beta_{12}=1$, the two Kramers doublets 
reduce to $|\pm\frac{1}{2}\rangle$ and $|\pm\frac{3}{2}\rangle$, 
which are belonging to $\Gamma_6$ and $\Gamma_7$ irreducible representations 
in D$_{\rm 4h}$ point group. In addition the operator $J^{\rm z}$ in 
$j$=3/2 quartet system is the same as the operator given in Eq.~(\ref{jz}) 
with $\alpha_{12}=\beta_{12}=1$. Therefore, when a magnetic field is applied 
in [001] direction, the present system with $\alpha_{12}=\beta_{12}=1$ 
is mapped to $j$=3/2 quartet system. However, such mapping is not applicable 
in magnetic field perpendicular along [001], due to differences of 
matrix elements of $J^{\rm x}$ and $J^{\rm y}$ between $j$=3/2 quartet system 
and the present system with $\alpha_{12}=\beta_{12}=1$.

\subsection{Inter-site term}
In the present case, we consider that $f$-hole localizes 
at each Yb-site. From the simplification mentioned above, 
we have one $\Gamma_6$ doublet and 
one $\Gamma_7$ doublet. 
Even in the simplification, there are 15 kinds of multipoles 
at each Yb-site. In order to describe the multipoles, 
we introduce bases of multipoles $\phi^{\Gamma}_{n}$ 
belonging to $\Gamma$-irreducible representation in D$_{\rm 4h}$ point group. 
In Table. \ref{d4h}, we classify $\phi^{\Gamma}_{n}$ 
according to irreducible representations in D$_{\rm 4h}$ point group, 
where in addition to $S^{\alpha}_{\tau\tau'}$, 
we use a $f$-charge operator 
\begin{eqnarray}
  \rho_{\tau\tau'}=\frac{1}{2}\sum_{\eta}
     f_{\tau\eta}^{\dagger}f_{\tau'\eta}.
\end{eqnarray} 
Since classification of multipoles up to octupole is also shown 
in Table. \ref{d4h}, 
correspondence between multipole and $\phi^{\Gamma}_{n}$ will be clear. 
For example, the x-component of dipole moment $J^{\rm x}$ is described by 
a linear combination of $\phi^{\Gamma_{5}^{-}}_{nx}$, which is 
consistent with Eq.(\ref{jx}).

Now, considering that metallic behavior has been observed in YbRu$_2$Ge$_2$, 
effective RKKY interactions between the multipoles are present, which 
are derived through the Schrieffer-Wolff 
transformation from hybridization term between 4$f$- and conduction-electrons. 
Noting that the Yb-sites in the compound form 
body-centered tetragonal lattice, 
inter-layer interactions should favor ferro-type order, 
since antiferro-couplings would lead to frustration. 
In the following, we consider multipole ordering within a mean-field theory 
for Yb in the body centered tetragonal structure, 
assuming that the ordering takes place either at the zone center 
$\tilde{\bf q}=0$ (ferro) or at the zone boundary 
$\tilde{\bf q}$=($\pi$,$\pi$,0) (antiferro). 
If we consider only diagonal term of nearest-neighbour couplings, 
the inter-site term of Hamiltonian $H_{\rm int}$ is given by
\begin{eqnarray}
  H_{\rm int}&=&-\sum_{{\bf i}\neq{\bf j}}\sum_{\Gamma,n}
             J^{\Gamma}_{{\bf i}-{\bf j}n}
             \phi^{\Gamma}_{{\bf i}n}\phi^{\Gamma}_{{\bf j}n},\\
  &=&-\frac{1}{N_0}\sum_{\Gamma,n}\sum_{\bf q}
             J^{\Gamma}_{n}({\bf q})
             \phi^{\Gamma}_{n}({\bf -q})\phi^{\Gamma}_{n}({\bf q}),
\end{eqnarray}
with 
$\phi^{\Gamma}_{n}({\bf q})=\sum_{\bf i}e^{-{\rm i}{\bf q}\dot{\bf i}}
\phi^{\Gamma}_{{\bf i}n}$, 
where $J^{\Gamma}_{n}$ is a coupling constant between 
$\phi^{\Gamma}_{n}$, and $N_0$ is the number of Yb-sites in crystal. 
In Eq. (14), 
$J^{\Gamma}_{n}({\bf q})$ and $\phi^{\Gamma}_{n}({\bf q})$ are 
Fourier components of coupling constants and 
$\phi^{\Gamma}_{{\bf i}n}$, respectively.
For square lattice, $J^{\Gamma}_{n}({\bf q})$ is given by
\begin{eqnarray}
  J^{\Gamma}_{n}({\bf q})=2J^{\Gamma}_{n}(\cos{q_x}+\cos{q_y}).
\end{eqnarray}

\subsection{Resulting effective Hamiltonian}
In order to explain low temperature property of YbRu$_2$Ge$_2$, 
effective Hamiltonian used in the following is described by
\begin{eqnarray}
  &&H_{\rm eff}=H_{\rm CEF}+H_{\rm Z}+H_{\rm int}.
\end{eqnarray}
As we have already mentioned, 
we can determine $|\Delta_0|$ from the entropy. 
However, in Zeeman term $H_{\rm Z}$, there are two free parameters, 
which control the weight of $|\pm1/2\rangle$ and $|\pm3/2\rangle$ in 
$\Gamma_{6}^{(1)}$ and $\Gamma_{7}^{(1)}$ states, respectively. 
In addition, we have coupling constants $J^{\Gamma}_{n}$, 
which will be estimated in the following sections. Then we apply a mean-field approximation for the effective Hamiltonian to calculate thermodynamic quantities and the phase diagrams.

\begin{table}[t]
\caption{Classification of multipoles and 
relevant bases of the multipoles $\phi^{\Gamma}_{n}$ 
according to irreducible representations of D$_{\rm 4h}$ point group. 
The first column shows 
irreducible representation of D$_{\rm 4h}$. 
The second and third ones describe multipoles and 
$\phi^{\Gamma}_{n}$ belonging to $\Gamma$ irreducible representation, 
respectively. Here $J$, $O$, and $T$ in the second column are 
dipole, quadrupole, and octupole moments, respectively. 
The fourth column shows expression of corresponding 
local susceptibility of $\phi^{\Gamma}_{n}$ 
with splitting energy $\Delta$ 
from $\Gamma_{6}$ to $\Gamma_{7}$ states.
The superscript $\pm$ of irreducible representation 
expresses the parity with respect to time reversal. }
\begin{center}
\begin{tabular}{cccc}
$\Gamma$ (D$_{\rm 4h}$) & multipole & $\phi^{\Gamma}_{n}$ 
 & $\chi^{{\rm L}\Gamma}_{nn}$\\
\hline
$\Gamma_{1}^{+}$ 
  & $O_{2}^{0}$ 
   & $\phi^{\Gamma_{1}^{+}}$=
     $\frac{1}{\sqrt{2}}(\rho_{66}-\rho_{77})$
    & $\frac{1}{8T}(1-\tanh^{2}{\frac{\Delta}{2T}})$\\
\hline
$\Gamma_{3}^{+}$ 
  & $O_{2}^{2}$ 
   & $\phi^{\Gamma_{3}^{+}}$=
     $\frac{1}{\sqrt{2}}(\rho_{67}+\rho_{76})$
    & $\frac{1}{4\Delta}\tanh{\frac{\Delta}{2T}}$\\
\hline
$\Gamma_{4}^{+}$ 
  & $O_{xy}$ 
   & $\phi^{\Gamma_{4}^{+}}$=
     $\frac{\rm i}{\sqrt{2}}(S^{\rm z}_{67}-S^{\rm z}_{76})$
    & $\frac{1}{4\Delta}\tanh{\frac{\Delta}{2T}}$\\
\hline
$\Gamma_{5}^{+}$ 
  & $O_{yz}$ 
   & $\phi^{\Gamma_{5}^{+}}_{x}$=
     $\frac{\rm i}{\sqrt{2}}(S^{\rm x}_{67}-S^{\rm x}_{76})$
    & $\frac{1}{4\Delta}\tanh{\frac{\Delta}{2T}}$\\
 \cline{2-4}
  & $O_{zx}$ 
   & $\phi^{\Gamma_{5}^{+}}_{y}$=
     $\frac{\rm i}{\sqrt{2}}(S^{\rm y}_{67}-S^{\rm y}_{76})$
    & $\frac{1}{4\Delta}\tanh{\frac{\Delta}{2T}}$\\
\hline\hline
$\Gamma_{2}^{-}$ 
  & $J^{\rm z}$ 
   & $\phi^{\Gamma_{2}^{-}}_{1}$=$S^{\rm z}_{66}$
    & $\frac{1}{8T}(1+\tanh{\frac{\Delta}{2T}})$\\
  & $T_{\rm z}^{\alpha}$ 
   & $\phi^{\Gamma_{2}^{-}}_{2}$=$S^{\rm z}_{77}$
    & $\frac{1}{8T}(1-\tanh{\frac{\Delta}{2T}})$\\
\hline
$\Gamma_{3}^{-}$ 
  & $T_{xyz}$ 
   & $\phi^{\Gamma_{3}^{-}}$=
     $\frac{\rm i}{\sqrt{2}}(\rho_{67}-\rho_{76})$
    & $\frac{1}{4\Delta}\tanh{\frac{\Delta}{2T}}$\\
\hline
$\Gamma_{4}^{-}$ 
  & $T_{\rm z}^{\beta}$ 
   & $\phi^{\Gamma_{4}^{-}}$=
     $\frac{1}{\sqrt{2}}(S^{\rm z}_{67}+S^{\rm z}_{76})$
    & $\frac{1}{4\Delta}\tanh{\frac{\Delta}{2T}}$\\
\hline
$\Gamma_{5}^{-}$ 
  & $J^{\rm x}$ 
   & $\phi^{\Gamma_{5}^{-}}_{1x}$=$S^{\rm x}_{66}$
    & $\frac{1}{8T}(1+\tanh{\frac{\Delta}{2T}})$\\
  & $T_{\rm x}^{\alpha}$ 
   & $\phi^{\Gamma_{5}^{-}}_{2x}$=$S^{\rm x}_{77}$
    & $\frac{1}{8T}(1-\tanh{\frac{\Delta}{2T}})$\\
  & $T_{\rm x}^{\beta}$ 
   & $\phi^{\Gamma_{5}^{-}}_{3x}$=
     $\frac{1}{\sqrt{2}}(S^{\rm x}_{67}+S^{\rm x}_{76})$
    & $\frac{1}{4\Delta}\tanh{\frac{\Delta}{2T}}$\\
 \cline{2-4}
  & $J^{\rm y}$ 
   & $\phi^{\Gamma_{5}^{-}}_{1y}$=$S^{\rm y}_{66}$
    & $\frac{1}{8T}(1+\tanh{\frac{\Delta}{2T}})$\\
  & $T_{\rm y}^{\alpha}$ 
   & $\phi^{\Gamma_{5}^{-}}_{2y}$=$S^{\rm y}_{77}$
    & $\frac{1}{8T}(1-\tanh{\frac{\Delta}{2T}})$\\
  & $T_{\rm y}^{\beta}$ 
   & $\phi^{\Gamma_{5}^{-}}_{3y}$=
     $\frac{1}{\sqrt{2}}(S^{\rm y}_{67}+S^{\rm y}_{76})$
    & $\frac{1}{4\Delta}\tanh{\frac{\Delta}{2T}}$\\
\end{tabular}
\end{center}
\label{d4h}
\end{table}

\section{Analysis of Transition at $T_0$}
In this section, we develop a mean-field theory for the effective 
Hamiltonian to analyze non-magnetic transition at $T_0$. 
After comparing phase diagrams for 
all types of ferro- and antiferro-quadrupole ordering 
with experimental one, 
we propose a prefered type of quadrupole order in YbRu$_{2}$Ge$_{2}$.

\subsection{Mean-field approximation for quadrupolar order}
At first, we give mean-field Hamiltonian 
to determine transition line in $H$-$T$ phase diagram. 
We assume that multipole ordered state is specified by 
irreducible representation $\Gamma$ 
and ordering wave vector $\tilde{\bf q}$. 
From the effective model, we obtain easily the mean-field Hamiltonian
\begin{eqnarray}
  &&H_{\rm MF}=H_{\rm CEF}+H_{\rm Z}+\tilde{H}_{\rm int},\\
  &&\tilde{H}_{\rm int}=-\frac{1}{N_0}\sum_{\Gamma,n}\sum_{\bf q}
             J^{\Gamma}_{n}({\bf q})
           (2\langle\phi^{\Gamma}_{n}({\bf q})\rangle\phi^{\Gamma}_{n}({\bf q})
\nonumber\\
  &&\hspace{41mm}-\langle\phi^{\Gamma}_{n}({\bf q})\rangle^{2}),\label{itmf}
\end{eqnarray}
with
\begin{eqnarray}
  \langle\cdots\rangle=\frac{{\rm Tr}\hspace{1mm}e^{-H_{\rm MF}/T}\cdots}
                            {{\rm Tr}\hspace{1mm}e^{-H_{\rm MF}/T}}
\end{eqnarray}
where $T$ is the temperature. 

We consider three cases, 
(1) system in zero magnetic field, 
(2) system in magnetic field parallel to [001] direction, and (3) system 
in magnetic field parallel to [100] direction, whose point groups are 
D$_{\rm 4h}$, C$_{\rm 4v}$, and C$_{\rm 2v}$, respectively. 
In Table \ref{c4v} and \ref{c2v}, 
bases of multipoles are classified according to irreducible representations 
of C$_{\rm 4v}$ and C$_{\rm 2v}$ point group, respectively. 
In order to develop a general formalism, 
we call bases of multipoles belonging to $\Gamma$-irreducible representation generically $\psi^{\Gamma}_{n}$ in any point group G (D$_{\rm 4h}$, C$_{\rm 4v}$, and C$_{\rm 2v}$). For D$_{\rm 4h}$ point group, $\psi^{\Gamma}_{n}$ 
is equivalent to $\phi^{\Gamma}_{n}$ given in Table. \ref{d4h} . 
For non-zero field the corresponding lower symmetry point groups C$_{\rm 4v}$, and C$_{\rm 2v}$ have basis functions $\psi^{\Gamma}_{n}$  that may still be directly expressed in terms of the D$_{\rm 4h}$ basis functions $\phi^{\Gamma}_{n}$ as shown in Tables II and III. 
In the following, we discuss only disordered, ferro-, and antiferro-ordered 
states, which is reasonable since we restrict to nearest-neighbor interaction 
in $H_{\rm int}$. 
In the disordered state only the fully symmetric multipole $\psi^{\Gamma_{1}}_{n}$ (O$_2^0$ in zero field) has a non-zero expectation value $\langle\psi^{\Gamma_{1}}_{n}({\bf 0})\rangle$. It leads to the background temperature dependence of $\Gamma_6-\Gamma_7$ splitting as shown later.
In ferro-ordered state belonging to $\Gamma$-irreducible representation, 
$\langle\psi^{\Gamma}_{n}({\bf 0})\rangle$ have finite values in addition to 
$\langle\psi^{\Gamma_{1}}_{n}({\bf 0})\rangle$. In antiferro-ordered 
state belonging to $\Gamma$-irreducible representation, 
allowed expectation values are 
$\langle\psi^{\Gamma}_{n}(\tilde{\bf q})\rangle$ and 
$\langle\psi^{\Gamma_{1}}_{n}({\bf 0})\rangle$.

\begin{table}[tb]
\caption{Classification of $\phi^{\Gamma}_{n}$ 
according to irreducible representations of C$_{\rm 4v}$, 
which is point group of tetragonal system in magnetic field parallel to [001]. 
Here, we note that $\phi^{\Gamma}_{n}$ are bases of multipoles 
belonging to $\Gamma$ irreducible representation in D$_{\rm 4h}$ 
point group. 
The first column shows 
irreducible representation of C$_{\rm 4v}$. 
The second, third, and fourth ones describe 
quadrupole, $\phi^{\Gamma^{+}}_{n}$, and 
$\phi^{\Gamma^{-}}_{n}$ belonging to each irreducible representation 
in C$_{\rm 4v}$, respectively. }
\begin{center}
\begin{tabular}{cccc}
$\Gamma$ (C$_{\rm 4v}$) & $O$ 
                        & $\phi^{\Gamma^{+}}_{n}$ (even) 
                        & $\phi^{\Gamma^{-}}_{n}$ (odd)\\
\hline
$\Gamma_{1}$ & $O_2^0$ 
   & $\phi^{\Gamma_{1}^{+}}$=
     $\frac{1}{\sqrt{2}}(\rho_{66}-\rho_{77})$
    & $\phi^{\Gamma_{2}^{-}}_{1}$=$S^{\rm z}_{66}$\\
  &&& $\phi^{\Gamma_{2}^{-}}_{2}$=$S^{\rm z}_{77}$\\
\hline
$\Gamma_{3}$ & $O_2^2$ 
   & $\phi^{\Gamma_{3}^{+}}$=
     $\frac{1}{\sqrt{2}}(\rho_{67}+\rho_{76})$
    & $\phi^{\Gamma_{4}^{-}}$=
     $\frac{1}{\sqrt{2}}(S^{\rm z}_{67}+S^{\rm z}_{76})$\\
\hline
$\Gamma_{4}$ & $O_{\rm xy}$ 
   & $\phi^{\Gamma_{4}^{+}}$=
     $\frac{\rm i}{\sqrt{2}}(S^{\rm z}_{67}-S^{\rm z}_{76})$
    & $\phi^{\Gamma_{3}^{-}}$=
      $\frac{\rm i}{\sqrt{2}}(\rho_{67}-\rho_{76})$\\
\hline
$\Gamma_{5}$ & $O_{\rm yz}$ 
   & $\phi^{\Gamma_{5}^{+}}_{x}$=
     $\frac{\rm i}{\sqrt{2}}(S^{\rm x}_{67}-S^{\rm x}_{76})$
    & $\phi^{\Gamma_{5}^{-}}_{1y}$=$S^{\rm y}_{66}$\\
  &&& $\phi^{\Gamma_{5}^{-}}_{2y}$=$S^{\rm y}_{77}$\\
  &&& $\phi^{\Gamma_{5}^{-}}_{3y}$=
     $\frac{1}{\sqrt{2}}(S^{\rm y}_{67}+S^{\rm y}_{76})$\\
 \cline{2-4}
  & $O_{\rm zx}$ 
   & $\phi^{\Gamma_{5}^{+}}_{y}$=
     $\frac{\rm i}{\sqrt{2}}(S^{\rm y}_{67}-S^{\rm y}_{76})$
    & $\phi^{\Gamma_{5}^{-}}_{1x}$=$S^{\rm x}_{66}$\\
  &&& $\phi^{\Gamma_{5}^{-}}_{2x}$=$S^{\rm x}_{77}$\\
  &&& $\phi^{\Gamma_{5}^{-}}_{3x}$=
     $\frac{1}{\sqrt{2}}(S^{\rm x}_{67}+S^{\rm x}_{76})$\\
\end{tabular}
\end{center}
\label{c4v}
\end{table}

\begin{table}[tb]
\caption{Classification of $\phi^{\Gamma}_{n}$ 
according to irreducible representations of C$_{\rm 2v}$, 
which is point group of tetragonal system in magnetic field parallel to [100]. 
Here, we note that $\phi^{\Gamma}_{n}$ are bases of multipoles 
belonging to $\Gamma$ irreducible representation in D$_{\rm 4h}$ 
point group. 
The first column shows 
irreducible representation of C$_{\rm 2v}$. 
The second, third, and fourth ones describe 
quadrupole, $\phi^{\Gamma^{+}}_{n}$, and 
$\phi^{\Gamma^{-}}_{n}$ belonging to each irreducible representation 
in C$_{\rm 2v}$, respectively. }
\begin{center}
\begin{tabular}{cccc}
$\Gamma$ (C$_{\rm 2v}$) & $O$ 
                        & $\phi^{\Gamma^{+}}_{n}$ (even) 
                         & $\phi^{\Gamma^{-}}_{n}$ (odd)\\
\hline
$\Gamma_{1}$ & $O_2^0$ 
  & $\phi^{\Gamma_{1}^{+}}$=
    $\frac{1}{\sqrt{2}}(\rho_{66}-\rho_{77})$
   & $\phi^{\Gamma_{5}^{-}}_{1x}$=$S^{\rm x}_{66}$\\
 & $O_2^2$ 
  & $\phi^{\Gamma_{3}^{+}}$=
    $\frac{1}{\sqrt{2}}(\rho_{67}+\rho_{76})$
   & $\phi^{\Gamma_{5}^{-}}_{2x}$=$S^{\rm x}_{77}$\\
 &&& $\phi^{\Gamma_{5}^{-}}_{3x}$=
     $\frac{1}{\sqrt{2}}(S^{\rm x}_{67}+S^{\rm x}_{76})$\\
\hline
$\Gamma_{2}$ & $O_{\rm yz}$ 
  & $\phi^{\Gamma_{5}^{+}}_{x}$=
    $\frac{\rm i}{\sqrt{2}}(S^{\rm x}_{67}-S^{\rm x}_{76})$
   & $\phi^{\Gamma_{3}^{-}}$=
     $\frac{\rm i}{\sqrt{2}}(\rho_{67}-\rho_{76})$\\
\hline
$\Gamma_{3}$ & $O_{\rm xy}$ 
  & $\phi^{\Gamma_{4}^{+}}$=
    $\frac{\rm i}{\sqrt{2}}(S^{\rm z}_{67}-S^{\rm z}_{76})$
   & $\phi^{\Gamma_{5}^{-}}_{1y}$=$S^{\rm y}_{66}$\\
 &&& $\phi^{\Gamma_{5}^{-}}_{2y}$=$S^{\rm y}_{77}$\\
 &&& $\phi^{\Gamma_{5}^{-}}_{3y}$=
     $\frac{1}{\sqrt{2}}(S^{\rm y}_{67}+S^{\rm y}_{76})$\\
\hline
$\Gamma_{4}$ & $O_{\rm zx}$ 
  & $\phi^{\Gamma_{5}^{+}}_{y}$=
    $\frac{\rm i}{\sqrt{2}}(S^{\rm y}_{67}-S^{\rm y}_{76})$
   & $\phi^{\Gamma_{2}^{-}}_{1}$=$S^{\rm z}_{66}$\\
 &&& $\phi^{\Gamma_{2}^{-}}_{2}$=$S^{\rm z}_{77}$\\
 &&& $\phi^{\Gamma_{4}^{-}}$=
     $\frac{1}{\sqrt{2}}(S^{\rm z}_{67}+S^{\rm z}_{76})$\\
\end{tabular}
\end{center}
\label{c2v}
\end{table}

For transition from disordered state to either ferro- or 
antiferro-multipole ordered state, 
we usually consider first- and second-order transitions. 
In a ferro- (or antiferro-) multipole ordered state 
belonging to $\Gamma$ irreducible representation in point group, 
expectation values of 
multipole moments $\langle\psi^{\Gamma}_{n}({\bf 0})\rangle$ 
(or $\langle\psi^{\Gamma}_{n}(\tilde{\bf q})\rangle$) 
have non-zero values. 
In general, 
free energy of ordered state is lower than that of disordered state 
below the transition temperature.
The explicit expression of free energy in mean field approximation has already been 
given by Shiina {\it{et al.}} \cite{Shiina}. 
In particular, if the transition is of second-order, 
all order parameters continuously reduce to zero 
on approaching the transition point. 
We use linearized mean-field equation 
to determine the second-order transition point, 
by expanding the partition function with respect to 
$\tilde{H}_{\rm int}$ given in Eq. (\ref{itmf}). 
For transition to ordered state specified by irreducible representation 
$\Gamma$ of point group and ordering wave vector $\tilde{\bf q}$ it is given by
\begin{eqnarray}
  \langle\psi^{\Gamma}_{n}(\tilde{\bf q})\rangle
  =\sum_{m}\chi^{{\rm L}\Gamma}_{nm}
   2J^{\Gamma}_{m}(\tilde{\bf q})
   \langle\psi^{\Gamma}_{m}(\tilde{\bf q})\rangle.\label{lmfe}
\end{eqnarray}
with local susceptibility $\chi^{{\rm L}\Gamma}_{nn'}$ defined by
\begin{eqnarray}
  \chi^{{\rm L}\Gamma}_{nn'}
  =\int^{1/T}_{0}d\tau\hspace{1mm}
   \langle\psi^{\Gamma}_{{\bf i}n}(\tau)
                   \psi^{\Gamma}_{{\bf i}n'}\rangle
   -\delta_{\Gamma=\Gamma_{1}}
    \frac{\langle\psi^{\Gamma_{1}}_{{\bf i}n}\rangle
          \langle\psi^{\Gamma_{1}}_{{\bf i}n'}\rangle}{T},\label{chil}
\end{eqnarray}
where $\tau$ is the imaginary time coordinate. 
The $\chi^{{\rm L}\Gamma}_{nn'}$ 
are calculated in the limit of 
$\langle\psi^{\Gamma}_{n}(\tilde{\bf q})\rangle\rightarrow$ 0. 
The linearized mean-field equation has non-trivial solution 
only when the maximum eigenvalue of 
2$\hat{\chi}^{{\rm L}\Gamma}J^{\Gamma}_{m}(\tilde{\bf q})$ becomes unity. 
We note that the transition becomes second-order, 
only if the transition point determined by linearized mean-field 
equation (\ref{lmfe}) is the same point 
where the free energies of the two states are equal.

\subsection{Estimation of coupling constants}
In our mean-field Hamiltonian, there are various parameters $\Delta_0$, $\alpha_{12}$, $\beta_{12}$ and $J^{\Gamma}_{n}$. 
At first, we consider disordered state in zero magnetic field. 
In this case, only allowed multipole is the uniform component of 
$\langle\phi^{\Gamma_{1}^{+}}\rangle$. 
Then, $\Delta_0$ is renormalized as 
\begin{eqnarray}
  \Delta
  =\Delta_{0}+\frac{1}{2}J^{\Gamma_{1}^{+}}({\bf 0})
   \tanh{\frac{\Delta}{2T}}.\label{delta}
\end{eqnarray}
For estimation of $\Delta_{0}$ and 
$J^{\Gamma_{1}^{+}}({\bf 0})$=$4J^{\Gamma_{1}^{+}}$, 
we assume $\Delta(T=12K)=$12K and $\Delta_{0}=$8K, which gives 
$\Delta(T=T_{0}=10.2K)=$12.8K and $J^{\Gamma_{1}^{+}}=$4.3K. 
By choosing these parameter values, we obtain reasonable temperature 
dependence of entropy around $T_0$. 
In the following calculation, we fix these values of $\Delta_{0}$ and 
$J^{\Gamma_{1}^{+}}$. 

Now we determine the values of $\alpha_{12}$ and $\beta_{12}$. 
These affect the magnetic anisotropy of the uniform susceptibility. 
Experimentally, 
the uniform susceptibility in magnetic field 
perpendicular to [001] is quite large compared to that in magnetic field 
parallel to [001]. 
The magnetic anisotropy is successfully reproduced 
by using the following lower two CEF states: 
The ground CEF state has almost pure $|\mp 1/2\rangle$ character 
belonging to $\Gamma_6$-irreducible representation, 
while the first-excited CEF state 
is almost pure $|\pm 3/2\rangle$ 
belonging to $\Gamma_7$-irreducible representation. 
Then the lower two CEF states are described by $\alpha_{12}\approx 1$ 
and $\beta_{12}\approx 1$, while $\alpha_{11}\approx\alpha_{22}\approx 0$
\cite{Geibel1}. 
With these values of $\alpha_{12}$ and $\beta_{12}$, 
contribution to $J^{\rm z}$ from the first excited $\Gamma_7$ state 
cancels out contribution from the ground $\Gamma_6$ state, 
while magnitude of $g$-factor $c_{66}^{\rm x}$ 
coming from the ground $\Gamma_6$ state in magnetic field 
parallel to [100] is four times larger than $c_{66}^{\rm z}$. 
Therefore, this CEF level scheme seems to be consistent with 
the observed magnetic anisotropy of the uniform susceptibility. 
In the following, we study transitions in the compound 
with these parameter values. 

Furthermore, we have many interaction parameters in the inter-site term. 
Because the highest transition takes place at $T_0$, 
each multipole interaction has an upper limit. 
We estimate the upper limit of each coupling constant, 
by assuming that the transition to each ordered state 
takes place at $T_0$ in zero field. 
Since we use the linearized mean-field equation (\ref{lmfe}), 
we need to evaluate $\chi^{{\rm L}\Gamma}_{nn'}$. 
In D$_{\rm 4h}$ point group, the matrix $\hat{\chi}^{{\rm L}\Gamma}$ 
has diagonal form for each $\Gamma$. 
Each eigenvalue of 
$n$-th component in $\Gamma$ is given by 
\begin{eqnarray}
  \lambda^{\Gamma}_{n}=2J^{\Gamma}_{n}(\tilde{\bf q})\chi^{{\rm L}\Gamma}_{nn}
\end{eqnarray}
where expressions for $\chi^{{\rm L}\Gamma}_{nn}$ are summarized in 4th 
column in Table. \ref{d4h}. 
Specifically, for each of the three two-dimensional representations, 
($J^{\rm x}$, $J^{\rm y}$), ($T^{\alpha}_{\rm x}$, $T^{\alpha}_{\rm y}$), 
and ($T^{\beta}_{\rm x}$, $T^{\beta}_{\rm y}$) in Table I, 
we assume that 
the upper limit of coupling constant is independent on $n$. 
Together with one-dimensional representation, the upper limit 
of coupling constant for $\Gamma$ irreducible representation 
$J^{\Gamma}_{\rm c}(\tilde{\bf q})$ is then given by
\begin{eqnarray}
  2J^{\Gamma}_{\rm c}(\tilde{\bf q})
   =\frac{1}{{\rm max}\hspace{1mm}\chi^{{\rm L}\Gamma}_{nn}(T=T_{0})}. 
\end{eqnarray}
In Table \ref{parameter}, by using 
the splitting energy $\Delta$ determined above, 
we list the calculated values of 
$1/8{\rm max}\hspace{1mm}\chi^{{\rm L}\Gamma}_{nn}(T=T_{0})$, 
which is equal to upper limit value of $J^{\Gamma}$ with factor 1/2$z$=1/8, 
where $z$ is the square lattice coordination number. 
If we use corresponding upper limit 
for every coupling constant, 
all transition temperatures to ordered states would become 
degenerate at $T_0$. 
In calculating the transition to an ordered state 
with a primary order parameter 
$\phi^{\Gamma}_{n}$, we assume that only the coupling constant 
for $\phi^{\Gamma}_{n}$ is at the upper limit. All others are 
reduced by a factor $\alpha_{\Gamma'}$ ($\Gamma'\neq\Gamma$) 
\begin{eqnarray}
  &&J^{\Gamma}_{n}(\tilde{\bf q})=J^{\Gamma}_{\rm c}(\tilde{\bf q}),\nonumber\\
  &&J^{\Gamma'}_{n}(\tilde{\bf q})
   =\alpha_{\Gamma'}J^{\Gamma'}_{\rm c}(\tilde{\bf q})
    \hspace{10mm}{\rm for}\hspace{2mm}\Gamma'\neq\Gamma
\end{eqnarray}
where $0<\alpha_{\Gamma'}<1$ is a ratio of actual coupling constant 
$J^{\Gamma}$ to the critical value $J^{\Gamma}_{\rm c}$. 
Present available experimental data do not allow a unique determination 
of coupling constants. Only that  for the primary ferroquadrupole order parameter may be fixed  by the transition temperature T$_0$.  In order to obtain a low-field behaviour of  $T_{0}(h)$ which is insensitive to the details of the model we assume  $\alpha_{\Gamma'}$=0.5
for the remaining coupling strengths.


\begin{table}[t]
\caption{Coupling constants used for analysis of YbRu$_2$Ge$_2$. 
The first column shows 
irreducible representation of D$_{\rm 4h}$. 
The second column shows upper limits of magnitudes of coupling constants
for having $\phi^{\Gamma}_{n}$ order parameter at T$_0$.
The third column shows coupling constants used in the model calculation 
of Sec. V.  Here plus/minus sign for  $J^{\Gamma}$ denotes ferro-/antiferro 
coupling. 
The value in parentheses 
exhibits reduced coupling constant between $\phi^{\Gamma_{5}^{-}}_{3y}$.}
\begin{center}
\begin{tabular}{cccc}
$\Gamma$ (D$_{\rm 4h}$) 
  & 1/8${\rm max}\hspace{1mm}\chi^{{\rm L}\Gamma}_{nn}(T=T_{0})$ [K] 
    & $J^{\Gamma}$ [K]\\
\hline
$\Gamma_{1}^{+}$ 
  & 14.8 & 4.3\\
\hline
$\Gamma_{3}^{+}$ 
  & 11.5 & 11.5\\
\hline
$\Gamma_{4}^{+}$ 
  & 11.5 & -3.0\\
\hline
$\Gamma_{5}^{+}$ 
  & 11.5 & -3.0\\
\hline\hline
$\Gamma_{2}^{-}$ 
  & 6.6 & 1.7\\
\hline
$\Gamma_{3}^{-}$ 
  & 11.5 & -3.0\\
\hline
$\Gamma_{4}^{-}$ 
  & 11.5 & -3.0\\
\hline
$\Gamma_{5}^{-}$ 
  & 6.6 & -3.4 (-2.1)\\
\end{tabular}
\end{center}
\label{parameter}
\end{table}

\subsection{Order parameter candidates for transition at $T_0$}
Before we discuss candidates for transition at $T_0$, 
we note a property of lower two CEF states 
for magnetic field parallel to [001]. 
As we have mentioned in previous subsection, the ground CEF state 
belonging to $\Gamma_6$ is lower than the first-excited CEF state 
belonging to $\Gamma_7$ by 12.8K at $T_0$ in zero magnetic field. 
Since the magnetic field parallel to [001] does not break 4-fold symmetry, 
we still can distinguish the $\Gamma_6$ and $\Gamma_7$ states. 
In this case their energies are easily obtained as 
$-\frac{\Delta}{2}\mp g_{J}\mu_{\rm B}c^{\rm z}_{66}h$ 
for the $\Gamma_6$ Kramers doublet
and $\frac{\Delta}{2}\mp g_{J}\mu_{\rm B}c^{\rm z}_{77}h$ 
for the $\Gamma_7$ state. 
There are two remarkable points in the present level scheme; 
(1) $c^{\rm z}_{66}$ has negative sign while $c^{\rm z}_{77}$ 
is positive, and (2) magnitude of $c^{\rm z}_{77}$ is three times 
larger than that of $c^{\rm z}_{66}$. 
From these facts we can expect a level crossing in the high field region of the disordered phase, where the CEF ground state changes 
from $\Gamma_6$ in low field to $\Gamma_7$ in high field\cite{comment1}. 

In the following  we discuss the transition line of $T_{0}(h)$ 
in magnetic field parallel to [001]. 
From experimental result, the specific heat data shows that 
the transition temperature of the non-magnetic phase decreases 
with increasing magnetic field. 
As mentioned before, 
we assume that the non-magnetic transition is obtained by ordering 
of quadrupole moment. 
However, we do not know the character of the quadrupole moment 
among the possible $O_{2}^{0}$, $O_{2}^{2}$, $O_{xy}$, and ($O_{yz}$, $O_{zx}$) cases,  and even whether the ordering is of ferro- or antiferro-type. 
In order to identify the type of quadrupole ordering, 
we calculate transition lines for all kinds of quadrupole orderings 
to compare with the experimental one. 
Firstly, since coupling constant between $O_{2}^{0}$ is ferro-like 
as estimated in the previous subsection, 
and ferro-component of $O_{2}^{0}$ is allowed even in the disordered 
state belonging to $\Gamma_1$ in C$_{\rm 4v}$ point group 
(Table \ref{c4v}), we exclude any ordering of $O_{2}^{0}$ 
from candidates of the transition at $T_0$. 
It does not break any symmetry and hence cannot lead to 
a specific heat jump. 
Secondly, we show transition lines for other types of quadrupole orderings 
in Fig. 1, where for coupling constants other 
than the primary order parameter, 
$\alpha_{\Gamma'}$=0.5 is chosen. 
Considering the behavior of experimental specific heat, 
ferro-$O_{2}^{2}$ and ferro-$O_{xy}$ orderings 
are consistent with the experiment among all types of quadrupole 
orderings. 
From calculated results we conclude that transition to ferro-$O_{2}^{2}$ 
(or ferro-$O_{xy}$) ordering state is of second-order 
in low magnetic field region. 
In order to explain a decrease of transition temperature 
in the magnetic field, 
we show schematic view of magnetic field dependence of level scheme 
with possible transitions for $O_{2}^{2}$ and $O_{xy}$ in Fig. 2. 
In both $O_{2}^{2}$ and $O_{xy}$ ordering states, 
the ground CEF state $|\Gamma_{6}+\rangle$ couples only with 
highest state $|\Gamma_{7}+\rangle$, therefore, 
$O_{2}^{2}$ and $O_{xy}$ order are of induced type  
because the quadrupole expectation value in the ground state vanishes. 
In addition, splitting energy between these states increases 
with increasing magnetic field. 
Then, since the energy denominator of the dominant term 
of corresponding local susceptibility 
Eq. (\ref{chil}) increases, the second-order transition temperature 
decreases with increasing magnetic field. 
On the other hand, for ordering of $O_{yz}$ or $O_{zx}$, 
transition temperature of these quadrupole orderings 
increases with increasing magnetic field, 
since splitting energy between the ground CEF state 
and excited state $|\Gamma_{7}-\rangle$ decreases. 

We now consider behavior in magnetic field perpendicular to [001]. 
From specific heat experiment, it is shown that the anomaly of specific heat 
jump broadens with increasing magnetic field. 
This means that 
the transition at $T_0$ in zero magnetic field reduces to 
a crossover in magnetic field perpendicular to [001]. 
We note that crossover does not break any symmetry. 
Instead of transition temperature $T_0$, we define 
a characteristic temperature $T^*$ as an inflection point of 
the specific heat divided by temperature. 
Let us assume antiferro-quadrupole ordering, 
which breaks at least translational symmetry. 
Since this would cause a specific heat jump, 
we exclude antiferro-quadrupole ordered state 
from the scenario of crossover. 
On the other hand, ferro-quadrupole ordering in general breaks 
local rotational symmetry, although the translational symmetry 
is preserved. 
However, ferro-quadrupole ordering belonging to 
$\Gamma_1$-irreducible representation 
neither breaks translational symmetry nor 
rotational symmetry. 
For the system in magnetic field parallel to [100], 
$O_{2}^{0}$ and $O_{2}^{2}$ belong to 
$\Gamma_1$-irreducible representation, 
given in Table \ref{c2v}. 
We note again that ferro-$O_{2}^{0}$ ordering 
does not break any symmetry for any field direction and therefore
is excluded.
Thus, we have one scenario that $T^*$ will be a crossover temperature 
from usual high temperature region to low temperature region 
where a considerable ferro-$O_{2}^{2}$ component appears. 

We now consider magnetic field along [110] direction, 
which is also perpendicular to tetragonal c axis. 
If we repeat similar discussion as for  [100] direction, the only possible ordering is 
ferro-$O_{xy}$  belonging to $\Gamma_1$ irreducible representation 
in corresponding point group 
(see Table V), instead of ferro-$O_{2}^{2}$. 
Unfortunately, we cannot distinguish ferro-$O_{2}^{2}$ 
from ferro-$O_{xy}$ ordering, 
because the anisotropy in ab plane is not clarified below $T_0$ 
from the present experimental data. 
Therefore, in order to distinguish the type of quadrupole ordering 
in YbRu$_{2}$Ge$_{2}$, determination of different thermodynamical 
properties for two magnetic field directions, [001] and [110], 
is required.  In particlular the possible lattice distortion 
induced by the quadrupole ordering should be investigated.
Furthermore, determination of elastic constant softening would be desirable. 

In addition, we consider difference of non-magnetic multipole moments 
between two $\Gamma_6$ (two $\Gamma_7$) level scheme 
and the $\Gamma_6$-$\Gamma_7$ system. 
In case of $\Gamma_6$-$\Gamma_7$ level scheme, 
non-magnetic multipole moments are summarized in Table I, II, III, and V, 
where $O_{2}^{2}$ and $O_{xy}$ are described by linear combination 
of off-diagonal operators $\rho_{\tau\tau'}$ and $S^{z}_{\tau\tau'}$ 
with orbital indices $\tau\neq\tau'$, respectively. 
On the other hand, in the cases of two $\Gamma_6$ and two $\Gamma_7$ 
level schemes, corresponding orbital off-diagonal operators 
are not of quadrupole-type and 
belong to $\Gamma_1$ and $\Gamma_2$ 
irreducible representations in D$_{\rm 4h}$ and C$_{\rm 4v}$ point groups, respectively.
This is because associated Kramers doublets belong to 
the same irreducible representation 
in such level schemes. 
Furthermore, in C$_{\rm 2v}$ point group these operators belong to $\Gamma_1$ and $\Gamma_3$  irreducible representations, respectively.
Due to these differences of irreducible representations 
from those in the $\Gamma_6$-$\Gamma_7$ level scheme, 
no consistent explanation for transition at $T_0$ is obtained 
for level schemes with two $\Gamma_6$ or two $\Gamma_7$ doublets. 
Considering these arguments, 
only $\Gamma_6$-$\Gamma_7$ level scheme provides 
a reasonable explanation for the transition at $T_0$. 

Finally, we summarize this section by proposing candidates 
for quadrupolar state below $T_0$. 
In this subsection, we have calculated transition lines of 
quadrupole ordering in magnetic field parallel to [001], 
and discussed specific heat jump in magnetic field perpendicular to [001]. 
Due to these results obtained from different point of views, 
only two appropriate candidates remain. 
The reasonable candidates for non-magnetic transition at $T_0$ 
are ferro-$O_{2}^{2}$ and ferro-$O_{xy}$ order parameters. 
In addition, these order are of induced type in the magnetic field 
parallel to [001]. 
Furthermore, we stress that both ferro-$O_{2}^{2}$ and ferro-$O_{xy}$ 
orderings can appear only for the $\Gamma_6$-$\Gamma_7$ level scheme.

\begin{figure}
\resizebox{75mm}{!}
{\includegraphics{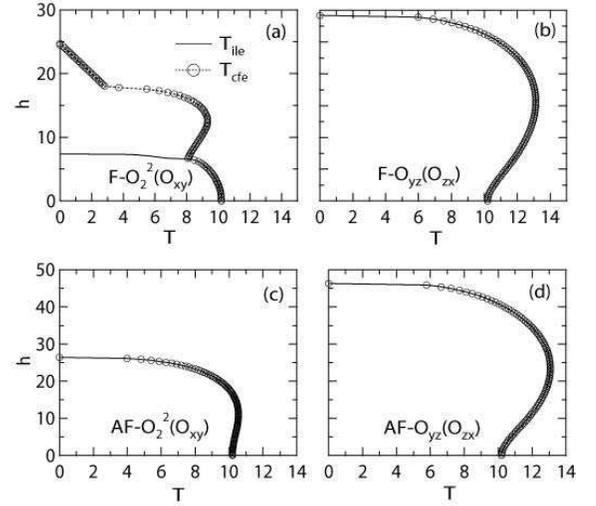}}
\caption{$h$-$T$ phase diagrams for quadrupole ordered states. 
The solid line determined by linearized mean-field equation. 
The open circle is given by comparison of free energies 
between disordered state and corresponding quadrupole ordering state. 
When these instability points are (not) the same, 
the transition is of second-order (first-order). 
It should be noted that 
$h_{\rm c}'(T_{0})=\infty$ in (a-d), 
which has been shown in Appendix B of Ref. 11.}
\end{figure}

\begin{figure}[t]
\resizebox{75mm}{!}
{\includegraphics{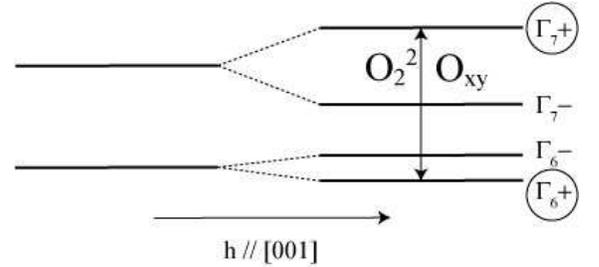}}
\caption{Schematic view of lower two CEF states in magnetic field 
parallel to [001]. By operators of $O_2^2$ and $O_{xy}$, 
which are equivalent to 
$\phi^{\Gamma_{3}^{+}}$=$\frac{1}{\sqrt{2}}(\rho_{67}+\rho_{76})$ and 
$\phi^{\Gamma_{4}^{+}}$=
$\frac{\rm i}{\sqrt{2}}(S^{\rm z}_{67}-S^{\rm z}_{76})$, respectively, 
the ground $\Gamma_6$ state couples only with the highest $\Gamma_7$ state 
in low magnetic field region. 
This leads to the induced type quadrupole order below $T_0$. 
The energy difference between these two states 
increases with increasing magnetic field. }
\end{figure}

\begin{table}[t]
\caption{Classification of $\phi^{\Gamma}_{n}$ 
according to irreducible representations of C$_{\rm 2v}$, 
which is point group of tetragonal system in magnetic field parallel to [110]. 
Here, we note that $\phi^{\Gamma}_{n}$ are bases of multipoles 
belonging to $\Gamma$ irreducible representation in D$_{\rm 4h}$ 
point group. 
The first column shows 
irreducible representation of C$_{\rm 2v}$. 
The second, third, and fourth ones describe 
quadrupole, $\phi^{\Gamma^{+}}_{n}$, and 
$\phi^{\Gamma^{-}}_{n}$ belonging to each irreducible representation 
in C$_{\rm 2v}$, respectively. }
\begin{center}
\begin{tabular}{cccc}
$\Gamma$ (C$_{\rm 2v}$) & quadrupole 
                        & $\phi^{\Gamma^{+}}_{n}$ (even) 
                         & $\phi^{\Gamma^{-}}_{n}$ (odd)\\
\hline
$\Gamma_{1}$ & $O_2^0$ 
  & $\phi^{\Gamma_{1}^{+}}$
   & $\phi^{\Gamma_{5}^{-}}_{1x}+\phi^{\Gamma_{5}^{-}}_{1y}$\\
 & $O_{\rm xy}$ 
  & $\phi^{\Gamma_{4}^{+}}$
   & $\phi^{\Gamma_{5}^{-}}_{2x}+\phi^{\Gamma_{5}^{-}}_{2y}$\\
 &&& $\phi^{\Gamma_{5}^{-}}_{3x}-\phi^{\Gamma_{5}^{-}}_{3y}$\\
\hline
$\Gamma_{2}$ & $O_{\rm yz}$-$O_{\rm zx}$ 
  & $\phi^{\Gamma_{5}^{+}}_{x}-\phi^{\Gamma_{5}^{+}}_{y}$
   & $\phi^{\Gamma_{4}^{-}}$\\
\hline
$\Gamma_{3}$ & $O_{\rm yz}$+$O_{\rm zx}$ 
  & $\phi^{\Gamma_{5}^{+}}_{x}+\phi^{\Gamma_{5}^{+}}_{y}$
   & $\phi^{\Gamma_{2}^{-}}_{1}$\\
 &&& $\phi^{\Gamma_{2}^{-}}_{2}$\\
 &&& $\phi^{\Gamma_{3}^{-}}$\\
\hline
$\Gamma_{4}$ & $O_2^2$ 
  & $\phi^{\Gamma_{3}^{+}}$
   & $\phi^{\Gamma_{5}^{-}}_{1x}-\phi^{\Gamma_{5}^{-}}_{1y}$\\
 &&& $\phi^{\Gamma_{5}^{-}}_{2x}-\phi^{\Gamma_{5}^{-}}_{2y}$\\
 &&& $\phi^{\Gamma_{5}^{-}}_{3x}+\phi^{\Gamma_{5}^{-}}_{3y}$\\
\end{tabular}
\end{center}
\end{table}

\section{Discussion of The Second Transition at $T_1$}
In the previous section, 
we have analyzed non-magnetic transition at $T_0$, 
and proposed either ferro-$O_{2}^{2}$ or ferro-$O_{xy}$ ordering 
in system with $\Gamma_6$-$\Gamma_7$ level scheme. 
In this section, we provide some proposals for the second transition 
at $T_1$, which is below $T_0$. 
%
%
At first, we summarize experimental results below $T_1$. 
Specific heat data show\cite{Jeevan} 
that a 2$^{nd}$ order phase transition 
appears at $T_1$. The transition temperature decreases with 
increasing magnetic field both parallel and perpendicular to [001]. 
Below $T_1$, uniform susceptibility data exhibit a clear difference between
the temperature dependence of $\chi_{\rm c}$ and $\chi_{\rm ab}$, 
which are the uniform susceptibilities 
in magnetic field parallel and perpendicular to [001], respectively. 
Here, $\chi_{\rm c}$ seems to be independent of temperature 
except for a slight decrease just below $T_1$, 
while $\chi_{\rm ab}$ shows a clear cusp at $T_1$ 
in weak magnetic field. 
From the difference of temperature dependence 
of  $\chi_{\rm c}$ and $\chi_{\rm ab}$ 
it has been proposed that below T$_1$ an antiferromagnetic state appears 
with staggered magnetic moment perpendicular to [001]. 
In addition magnetization for field perpendicular 
to [001] has shown a metamagnetic transition. This has been regarded 
as a spin-flop transition in the $ab$-plane. 

From a theoretical point of view, time reversal symmetry must be broken 
eventually below non-magnetic transition temperature $T_0$, 
in order to release remaining entropy $R\ln{2}$ 
of the Kramers doublet ground state. 
In this sense, the antiferromagnetic ordering is reasonable. 
However, in addition to dipole (magnetic) moments
the present model has octupole moments 
which also break time reversal symmetry.
Below the quadrupolar transition temperature $T_0$, 
the point group reduces from D$_{\rm 4h}$ to D$_{\rm 2h}$ 
in zero magnetic field, 
because both ferro-$O_{2}^{2}$ and ferro-$O_{xy}$ ordering 
break only four-fold rotational symmetry. 
For dipole and octupole moments, there are four one-dimensional 
irreducible representations in D$_{\rm 2h}$ point group. 
Among these four irreducible representations, three have 
respective component of dipole moment 
$J^{\alpha}$ in addition to two components of octupoles. 
On the other hand, the remaining irreducible representation has 
only one octupole component. 
Let us consider the uniform magnetic susceptibility 
in pure octupole ordered state. 
In very weak magnetic field, it is expected that 
the uniform susceptibility does not considerably 
decrease below the transition temperature of the octupole ordering, 
because there is no dipole order parameter in the state. 
Since this behavior is inconsistent with experimental data of $\chi_{\rm ab}$, 
we exclude the pure octupole from candidates for the ordered state 
below $T_1$. 
Therefore, even though we include octupole degrees of freedom, 
the state below $T_1$ seems to be inconsistent 
with pure octupole ordering, 
but rather must be an antiferromagnetic state with a considerable magnitude 
of the dipole moment. 

As candidate below $T_1$, an antiferromagnetic state will be 
reasonable from the above discussion of the susceptibility. 
However, the direction of the staggered moments in the ab plane  is still unclear. The direction depends on whether the
ordering quadrupole moment below $T_0$ is $O_{2}^{2}$ or $O_{xy}$ because their remaining symmetry axis is different. Let us consider ferro-$O_{2}^{2}$ 
order in magnetic field parallel to [100]. 
According to Table \ref{c2v} the
allowed directions of magnetic moments in the system are 
$J^{x}{\bf e}_{x}$, $J^{y}{\bf e}_{y}$, and $J^{z}{\bf e}_{z}$, 
where ${\bf e}_{x}$, ${\bf e}_{y}$, and ${\bf e}_{z}$ are unit vectors 
parallel to [100], [010], and [001] directions, respectively. 
Furthermore the direction of staggered moments 
in the antiferromagnetic state is perpendicular to [001] axis 
according to the experimental uniform susceptibility. 
Therefore, for magnetic transition within the $O_{2}^{2}$ phase, 
the dipole order parameter should be described by 
$J^{x}{\bf e}_{x}$ or $J^{y}{\bf e}_{y}$. 
Likewise within the $O_{xy}$ phase
in magnetic field parallel to [110], 
it should be $J^{x}{\bf e}_{x}\pm J^{y}{\bf e}_{y}$. 
This means that determination of staggered moment direction can distinguish 
between quadrupole order of ferro-$O_{2}^{2}$ or ferro-$O_{xy}$ type. 

Now we consider the metamagnetic transition, which has been observed 
in magnetic field perpendicular to [001], starting from
an antiferromagnetic state with planar staggered moment. 
In the present scenario, the antiferromagnetism is regarded 
to appear below ferro-$O_{2}^{2}$ or ferro-$O_{xy}$ ordering temperature. 
In these cases the point group 
reduces from D$_{\rm 4h}$ to D$_{\rm 2h}$ in zero magnetic field. 
Consequently two-dimensional irreducible representation in D$_{\rm 4h}$, 
which involves two planar components of magnetic moment, 
reduce to the direct sum of two one-dimensional 
irreducible representations, 
where each has 
one planar component of magnetic moment. 

Then it is expected that exchange coupling constants between planar components 
of magnetic moments depend on the in-plane direction. 
These effective coupling constants are due to RKKY mechanism, 
therefore their anisotropy will be induced by reconstruction of conduction
electron states in the ferro-quadrupolar ordered phase.
This exchange anisotropy induced by quadrupole order is the origin 
of the spin flop transition for $H\perp$[001]. 

\section{Mean-field analysis of both ordered phases}
From previous discussions, we have two candidates for successive 
transitions; one scenario is given by ferro-$O_{2}^{2}$ ordering 
for the first transition at $T_0$ before the second transition at $T_1$ 
to antiferromagnetism with staggered moment 
$J^{x}{\bf e}_{x}$ or $J^{y}{\bf e}_{y}$, 
and another is ferro-$O_{xy}$ ordering for transition at $T_0$ 
before transition at $T_1$ to antiferromagnetism with staggered moment 
$J^{x}{\bf e}_{x}\pm J^{y}{\bf e}_{y}$. 
In the following, we assume ferro-$O_{2}^{2}$ ordering for the transition 
at $T_0$, because we cannot distinguish these two possibilities
from available experimental data. 

\subsection{Re-estimation of coupling constants}
In the previous section, we have estimated parameter values of 
model Hamiltonian, such that magnetic anisotropy of uniform susceptibility 
and non-magnetic transition temperature are reasonably reproduced. 
However, in that stage, we have assumed that all signs of coupling 
constants are the same. 
In the present case, we are considering that the system shows 
ferro-quadrupole ordering before antiferromagnetic transition. 
Therefore, we should estimate coupling constants with assumption 
of ferro-quadrupole transition at $T_0$ and 
antiferromagnetic transition at $T_1$. 
Among these coupling constants, $J^{\Gamma_{1}^{+}}$ and $J^{\Gamma_{3}^{+}}$ 
are not changed from values used for Fig. 1(a). 
On the other hand, we choose value of $J^{\Gamma_{5}^{-}}$ 
as antiferromagnetic transition appears at $T_1$=6.5K in zero magnetic field. 
In addition, magnitudes of other coupling constants are chosen 
to be small with $\alpha^{\Gamma'}$=0.25 ($\Gamma'$=$\Gamma_{4}^{+}$, 
$\Gamma_{5}^{+}$, $\Gamma_{2}^{-}$, $\Gamma_{3}^{-}$, and $\Gamma_{4}^{-}$), 
so that unobserved phases are suppressed which would appear 
if $\alpha^{\Gamma'}$ were larger. 
Furthermore, considering that coupling constant between x-components 
of magnetic moment is different from correspondence between y-components 
of magnetic moment in ferro-$O_{2}^{2}$ ordering state, 
we will reduce magnitude of coupling constant 
between $\phi^{\Gamma_{5}^{-}}_{3y}$. 
In Table \ref{parameter}, 
we summarize the revised values of coupling constants. 

\subsection{Phase diagram}
At first, we construct $H$-$T$ phase diagram in magnetic field 
parallel to [001]. 
As we have mentioned in \S III.A, 
in order to find the transition line, we use two kinds of procedures; 
one is given by linearized mean-field equation, 
while the other is determined by 
comparison of free energies of different states. 
In Fig. 3, we show calculated $H$-$T$ phase diagram 
in magnetic field parallel to [001]. 
In the high field region of the phase diagram, 
level crossing between $\Gamma_6$ 
and $\Gamma_7$ states is obtained in terms of larger magnitude and 
negative sign of $g$-factor of the $\Gamma_7$ state. 
The transition at $T_0$ is of second-order, and 
$T_0$ decreases with increasing magnetic field. 
The transition at $T_1$ is second-order transition from ferro-$O_2^2$ 
phase to coexistent phase of ferro-$O_2^2$ and antiferromagnetism. 
Here, dashed line shows instability of paramagnetic state 
to ferro-$O_2^2$ state, while dotted line is instability line of 
paramagnetic state to antiferromagnetic state with staggered moment 
parallel to [100]. 
The dashed line is very different from instability line shown in Fig. 1, 
since coupling constant between field induced $\phi^{\Gamma_{4}^{-}}$ is 
changed from ferro-coupling to antiferro-coupling. 
Comparing these instability lines with the transition line 
to the coexistent phase, 
ferro-$O_2^2$ and antiferromagnetism are cooperative to each other 
and stabilize the coexistent phase.

\begin{figure}[t]
\resizebox{75mm}{!}
{\includegraphics{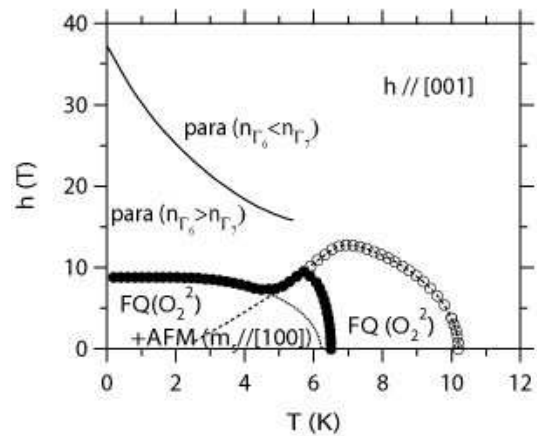}}
\caption{$h$-$T$ phase diagram in magnetic field parallel to [001]. 
The solid line shows level crossing temperature. 
The dashed and dotted lines are instability lines of disordered state 
to ferro-$O_2^2$ state and antiferromagnetic state with staggered 
moment parallel to [100], respectively. 
The open circle corresponds to transition point from disordered state 
to ferro-$O_2^2$ ordered state. 
The solid circle corresponds to transition point from paramagnetic state 
(including ferro-$O_2^2$ ordered state) to coexistent state of 
both ferro-$O_2^2$ moment and staggered magnetic moment parallel to [100]. 
These circles are determined by comparison of free energies. }
\end{figure}

In Fig. 4, we show calculated $H$-$T$ phase diagram 
in magnetic field parallel to [100]. 
In this figure, $T_0$ is not a transition temperature but the crossover 
temperature from usual paramagnetic phase in high temperature region 
to disordered phase with considerable ferro-$O_2^2$ moment 
in low temperature region, 
where the crossover temperature is determined by inflection point 
of temperature dependence of $C/T$. 
The crossover temperature increases with increasing magnetic field.
In order to consider low temperature state, 
we take into account reduced coupling constant between $J^{y}$, 
which is mentioned in \S.IV. 
By the anisotropic magnetic interaction, 
we have two antiferromagnetic phases. 
Here, low field phase has staggered moment parallel to the magnetic field, 
while high field phase has perpendicular component to the magnetic field. 
Therefore, spin-flop transition is obtained, 
where the transition between these phases is of first-order. 
If we use the same coupling constant as the one between $J^{x}$, 
low field antiferromagnetic phase disappears. 
In the inset of Fig. 4, metamagnetic transition corresponding 
to the spin-flop is obtained.

\begin{figure}[t]
\resizebox{75mm}{!}
{\includegraphics{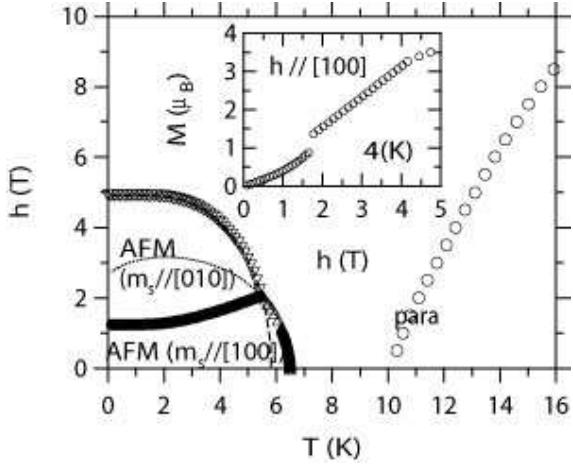}}
\caption{$h$-$T$ phase diagram in magnetic field parallel to [100]. 
The open circle shows crossover temperature determined by inflection 
point of temperature dependence of $C/T$. 
The dotted and dashed lines are instability lines of disordered state 
to antiferromagnetic state with staggered 
moment parallel to [100] and parallel to [010], respectively. 
The up-triangle and down-triangle correspond to transition point 
from disordered state to antiferromagnetic state with staggered moment 
parallel to [100] and parallel to [010], respectively, which are 
determined by comparison of free energies. 
Inset shows magnetization in field parallel to [100]. }
\end{figure}

\subsection{Specific heat}
In specific heat data, there are some anomalous features in the temperature 
and field dependences. In magnetic field parallel to [001], 
specific heat jumps are observed at $T_0$ and $T_1$, 
and these transition temperatures decrease 
with increasing magnetic field. 
In magnetic field perpendicular to [001], the anomaly at the non-magnetic 
transition reduces to a hump and broadens with increasing magnetic field. 
In order to compare with the experimental data, we calculate 
specific heat in magnetic field, based on the mean-field solution. 

In order to calculate specific heat, 
we first provide expression of internal energy $E_{\rm I}$, as follows:
\begin{eqnarray}
  E_{\rm I}=\sum_{\bf j}\sum_{n}
            \epsilon_{n}\langle\psi^{\Gamma_{1}}_{{\bf j}n}\rangle
           -\sum_{{\bf i}\neq{\bf j}}\sum_{\Gamma,n}
            J^{\Gamma}_{{\bf i}-{\bf j}n}
             \langle\psi^{\Gamma}_{{\bf i}n}\rangle
             \langle\psi^{\Gamma}_{{\bf j}n}\rangle,
\end{eqnarray}
where the first term of right hand side exhibits contributions 
from CEF term and Zeeman term. 
Then, specific heat is given by temperature derivative of 
the internal energy as
\begin{eqnarray}
  &&C=\sum_{\bf j}\sum_{n}\epsilon_{n}
    \frac{d\langle\psi^{\Gamma_{1}}_{{\bf j}n}\rangle}{dT}\nonumber\\
  &&\hspace{5mm}-\sum_{{\bf i}\neq{\bf j}}\sum_{\Gamma,n}
        2J^{\Gamma}_{{\bf i}-{\bf j}n}
        \langle\psi^{\Gamma}_{{\bf i}n}\rangle
        \frac{d\langle\psi^{\Gamma}_{{\bf j}n}\rangle}{dT},
\end{eqnarray}
with equation of 
temperature derivative of $\langle\psi^{\Gamma}_{{\bf j}n}\rangle$
\begin{eqnarray}
  &&\frac{d\langle\psi^{\Gamma}_{{\bf j}n}\rangle}{dT}
   =\sum_{n'}\frac{\epsilon_{n'}}{T}\chi^{\Gamma_{1}\Gamma}_{{\bf j}n'n}
    \nonumber\\
  &&\hspace{5mm}-\sum_{{\bf i}(\neq{\bf j})}\sum_{\Gamma',n'}
        2J^{\Gamma'}_{{\bf i}-{\bf j}n'}
        (\frac{\langle\psi^{\Gamma'}_{{\bf i}n'}\rangle}{T}
        -\frac{d\langle\psi^{\Gamma'}_{{\bf i}n'}\rangle}{dT})
        \chi^{\Gamma'\Gamma}_{{\bf j}n'n}.
\end{eqnarray}
where $\chi^{\Gamma'\Gamma}_{{\bf i}n'n}$ is given by 
the sublattice dependent expression
\begin{eqnarray}
  \chi^{\Gamma'\Gamma}_{{\bf i}n'n}
  =\int^{1/T}_{0}d\tau\hspace{1mm}
   \langle\psi^{\Gamma'}_{{\bf i}n'}(\tau)\psi^{\Gamma}_{{\bf i}n}\rangle
   -\frac{\langle\psi^{\Gamma'}_{{\bf i}n'}\rangle
          \langle\psi^{\Gamma}_{{\bf i}n}\rangle}{T}.
\end{eqnarray}

Using these expressions, we calculate the specific heat of the system 
in a magnetic field. 
In Fig. 5(a) and 5(b), we show temperature dependences of specific heat 
for field parallel to [001] and [100] directions, respectively. 
In Fig. 5(a), there are two specific heat jumps corresponding to 
transitions to ferro-$O_2^2$ state at $T_0$ 
and to coexistent state at $T_1$. 
In addition, field dependence of the specific heat is very weak 
in low field region. 
In Fig. 5(b), only specific heat jump is due to transition to 
antiferromagnetic state at $T_1$, while hump structure related to 
the crossover temperature at $T_0$ broadens 
with increasing magnetic field. 
This means that the ferro-$O_2^2$ moment is induced by magnetic field 
parallel to [100] as shown in Table \ref{c2v} and 
no symmetry breaking at $T_0$ takes place for $H>0$.

\begin{figure}[t]
\resizebox{75mm}{!}
{\includegraphics{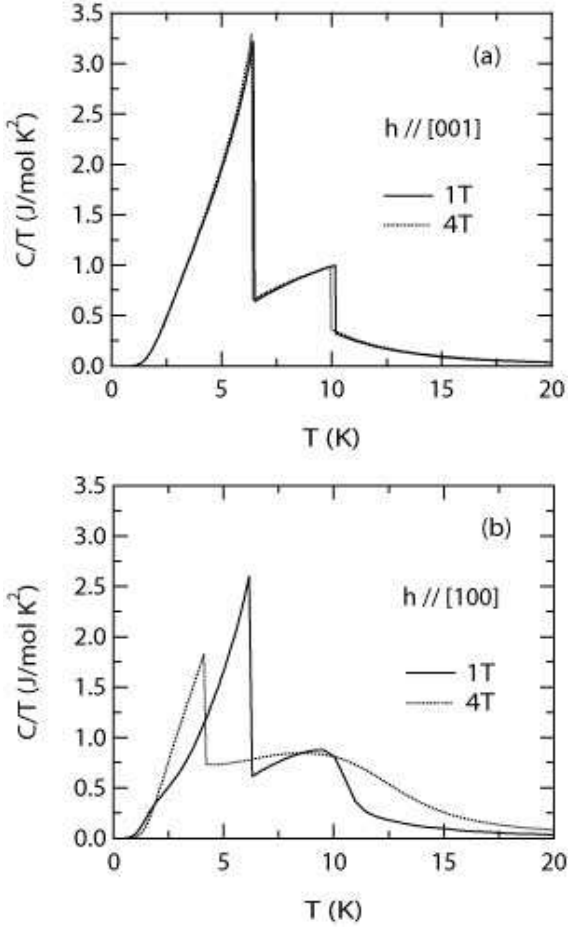}}
\caption{Temperature dependences of specific heat in magnetic field 
parallel to [001] (a) and parallel to [100] (b). 
The solid and dotted lines correspond to 1T and 4T, respectively. }
\end{figure}

\subsection{Uniform susceptibility}
The experimental data of uniform susceptibility exhibits characteristic 
properties. 
In the magnetic field perpendicular to [001], 
uniform susceptibility does not show anomaly at $T_0$, 
except for the slight increase of magnitude of the temperature derivative 
at $T_0$ in low field region. 
On the other hand, temperature dependence of 
uniform susceptibility in magnetic field parallel to [001] 
has a plateau-like behavior between $T_0$ and $T_1$. 
In addition, the temperature dependence is almost insensitive to 
the magnetic field up to 3T. In order to analyze uniform susceptibility, 
we give expressions of the quantity in magnetic field parallel to 
[001] and [100]. 
For direct comparison with experimental data in finite magnetic field, 
magnetization divided by magnitude of the field is used 
as uniform susceptibility, 
instead of the Kubo formula of susceptibility from linear response theory. 
Then, uniform susceptibility $\chi_{\rm c}$ 
in magnetic field parallel to [001] is given by
\begin{eqnarray}
  \chi_{\rm c}=\frac{g_{J}\mu_{\rm B}}{h}\langle J^{\rm z}\rangle,
\end{eqnarray}
while uniform susceptibility in magnetic field parallel to [100] 
is similarly obtained from
\begin{eqnarray}
  \chi_{\rm a}=\frac{g_{J}\mu_{\rm B}}{h}\langle J^{\rm x}\rangle,
\end{eqnarray}
where $J^{\rm z}$ and $J^{\rm x}$ are given by eqs. (6) and (7), 
respectively. 

In Fig. 6(a), we show calculated temperature dependence of 
uniform susceptibility in magnetic field parallel to [001]. 
Anomalies due to two transitions at $T_0$ and $T_1$ are obtained. 
Here, we note that plateau in temperature dependence between $T_0$ and $T_1$ 
is due to moderate coupling constant between octupoles 
$\phi^{\Gamma_{4}^{-}}$, which are induced by the magnetic field, 
while temperature independent behavior well below $T_1$ 
reflects existence of staggered moment perpendicular to the magnetic field. 
In addition, the uniform susceptibility hardly has field dependence 
at least in the low field region. 
In Fig. 6(b), we show temperature dependence of uniform susceptibility 
in magnetic field parallel to [100]. 
From the figure it shows slight upturn 
around crossover temperature $T_0$. 
In magnetic field of 1T, it shows a cusp at 
transition temperature to antiferromagnetism with staggered moment 
parallel to the magnetic field. 
With increase of magnetic field to 4T, the uniform susceptibility 
shows temperature independent behavior below transition temperature 
to antiferromagnetism with staggered moment perpendicular to 
the magnetic field. 
The different of behavior below antiferromagnetic transition temperature 
is due to the spin-flop transition, as shown in Fig. 4.

\begin{figure}[t]
\resizebox{75mm}{!}
{\includegraphics{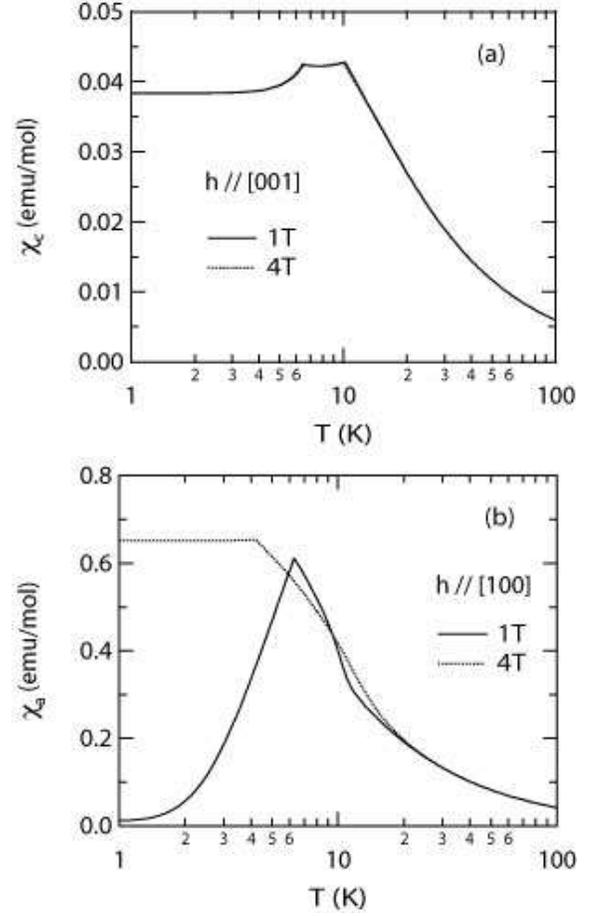}}
\caption{Temperature dependences of uniform susceptibilities 
in magnetic field parallel to [001] (a) and parallel to [100]. 
The solid and dotted lines correspond to 1T and 4T, respectively. }
\end{figure}

\section{discussion and summary}
Before we summarize, we would like to comment on a few points. 
In the previous section, we have calculated specific heat, 
uniform susceptibility, and phase diagram with assumption of 
ferro-$O_2^2$ ordering at $T_0$. 
Comparing our results with experimental data of YbRu$_2$Ge$_2$, 
the present model not only explains the experimental data 
qualitatively, but also gives quantitative agreement in specific heat jumps. 
Therefore, ferro-$O_2^2$ ordering at $T_0$ above antiferromagnetic 
transition temperature $T_1$ will be one of promising candidate for 
non-magnetic transition of YbRu$_2$Ge$_2$. 
However, we have proposed either ferro-$O_2^2$ or ferro-$O_{xy}$ ordering 
for the transition at $T_0$ in \S. III. 
Considering that both candidates are ferro-ordering states of 
quadrupoles, in order to identify the non-magnetic state 
among the two candidates, 
it will be useful to carry out ultrasonic and x-ray scattering experiments
, 
because ferro-quadrupole couples with lattice distortion. 
From this point of view, it is desirable to confirm 
the crystal structure below $T_0$. 

Related to property of non-magnetic ordering state, 
it has been recently reported that Ru-NQR spectrum may not be affected by 
the non-magnetic transition\cite{Mukuda}. 
With respect to this result, 
taking into account positions of Yb and Ru ions, 
if dominant quadrupole of Ru-nucleus is either $O_2^0$ or $O_2^2$, 
ferro-$O_{xy}$ ordering of Yb$^{3+}$ does not change Ru-NQR spectrum 
by the transition. 
On the other hand, 
if quadrupole of Ru-nucleus is $O_{xy}$, 
ferro-$O_2^2$ ordering does not change the spectrum by the transition. 
Furthermore, if quadrupole of Ru-nucleus is either $O_{yz}$ or $O_{zx}$, 
both ferro-$O_2^2$ ordering and ferro-$O_{xy}$ ordering 
do not change the spectrum. 
In order to identify the type of quadrupole ordering, 
it is required to clarify the quadrupole of Ru-nucleus. 

Furthermore, in recent uniform susceptibility data, it is shown that 
behavior of uniform susceptibility in magnetic field parallel to [100] 
is similar as that in magnetic field parallel to [110], 
and these uniform susceptibilities have finite value 
at very low temperature. 
Based on the present model with assumption of either ferro-$O_2^2$ 
or ferro-$O_{xy}$ ordering at $T_0$, one of these uniform susceptibility 
will vanish at 0K in small magnetic field region 
by development of staggered magnetic moment parallel to the magnetic field. 
Considering that finite values are observed for both uniform susceptibilities 
in very low temperature region, 
the ordering wave vector of magnetic moments will be
an incommensurate one. 
Therefore, it is desirable to carry out neutron scattering experiment 
to clarify the magnetic state of the compound. 

Finally, we comment on effect of multipolar fluctuations. 
The effect has been investigated in CeB$_6$ by Shiina\cite{Shiina2}. 
In the paper, it has been shown that the multipolar fluctuation 
are enhanced by approaching the system to the SU(4) symmetric limit. 
In YbRu$_2$Ge$_2$, it is considered that the tetragonal anisotropy 
like splitting energy $\Delta$ between $\Gamma_6$ and $\Gamma_7$ states 
breaks the SU(4) symmetry inherently. Therefore we do not expect 
that the multipolar fluctuations change qualitatively behaviors 
suggested by the mean-field theory. 

In summary, in order to explain properties of YbRu$_2$Ge$_2$, 
we have introduced a quasi-degenerate localized model 
consisting of CEF term, Zeeman term, and exchange term of multipoles. 
Classifying multipoles according to irreducible representations of 
corresponding point group, we have developed a mean-field theory 
for the model. 
Considering that the specific heat jump broadens with increasing 
magnetic field perpendicular to [001], 
we have proposed that for the non-magnetic transition, 
only ferro-$O_2^2$ and ferro-$O_{xy}$ orderings are possible candidates. 
Furthermore, it has been shown that these ferro-quadrupole orderings 
are only available and essentially of the induced type, 
when the lower two CEF states consist of 
one $\Gamma_6$ and one $\Gamma_7$ doublets in zero magnetic field. 
With assumption of ferro-$O_2^2$ ordering at $T_0$, 
we have calculated specific heat, uniform susceptibility, 
and phase diagram, 
where anisotropic exchange interaction between planar components of
magnetic moments 
is introduced as an effect of the ferro-quadrupole ordering. 
The calculated results have been shown to explain experimental 
data consistently. 
In order to clarify the property of YbRu$_2$Ge$_2$ completely 
and refine the set of coupling constants, 
it is desirable to carry out more detailed experiments 
like elastic constant measurements and neutron diffraction 
in applied magnetic field.


\section*{Acknowledgement}

The authors would like to thank C. Geibel and H. Mukuda 
for many valuable discussions with respect to experimental data. 

\vskip-0.5cm


\begin{thebibliography}{99}
\bibitem{comment1}
In Fig. 1(a), we can see the sign of the level cross 
around 20T. 
Here, we should note that this first-order transition is different 
from non-magnetic transition observed in YbRu$_2$Ge$_2$, 
because this transition is not expected 
in low-field region.
\end{thebibliography}
\end{document}